\documentclass[aps, prb, twocolumn, superscriptaddress, letterpaper, floatfix]{revtex4-2}

\usepackage{graphicx}
\usepackage{float}
\usepackage{amssymb}
\usepackage{amsmath}
\usepackage{txfonts}
\usepackage{bm}
\usepackage[dvipsnames]{xcolor}
\usepackage[colorlinks=true, allcolors=Mulberry]{hyperref}
\usepackage{orcidlink}

\graphicspath{{./}}
\newcommand{\SM}[1]{\textcolor{Mulberry}{#1}}

\usepackage{helvet}
\newcommand{\mysec}[1]{%
    \par\medskip
    \noindent\textbf{\fontfamily{phv}\selectfont{#1}}%
}

\begin{document}

\title{Scaling laws of entanglement entropy in fractal dimensions}

\author{Zheyu Cheng\orcidlink{0000-0001-5009-7929}}
\affiliation{Division of Physics and Applied Physics, School of Physical and Mathematical Sciences, Nanyang Technological University,
Singapore 637371, Singapore}

\author{Kailin Song}
\affiliation{Division of Physics and Applied Physics, School of Physical and Mathematical Sciences, Nanyang Technological University,
Singapore 637371, Singapore}

\author{Jian-Wei Liu}
\affiliation{Division of Physics and Applied Physics, School of Physical and Mathematical Sciences, Nanyang Technological University,
Singapore 637371, Singapore}

\author{Hanxi Sun}
\affiliation{Division of Physics and Applied Physics, School of Physical and Mathematical Sciences, Nanyang Technological University,
Singapore 637371, Singapore}

\author{Hongyu Zou}
\affiliation{School of Electrical and Electronic Engineering, Nanyang Technological University, Singapore 639798, Singapore}

\author{Yuanjin Zheng}
\email{yjzheng@ntu.edu.sg}
\affiliation{School of Electrical and Electronic Engineering, Nanyang Technological University, Singapore 639798, Singapore}

\author{Yihao Yang\orcidlink{0000-0002-9229-0054}}
\email{yangyihao@zju.edu.cn}
\affiliation{State Key Laboratory of Extreme Photonics and Instrumentation, ZJU-Hangzhou Global Scientific and Technological Innovation Center, Zhejiang University, Hangzhou, China}
\affiliation{International Joint Innovation Center, The Electromagnetics Academy at Zhejiang University, Zhejiang University, Haining, China}
\affiliation{Key Lab of Advanced Micro/Nano ElectronicDevices \& Smart Systems of Zhejiang, Jinhua Institute of Zhejiang University, Zhejiang University, Jinhua, China}
\affiliation{Shaoxing Institute of Zhejiang University, Zhejiang University, Shaoxing, China}

\author{Baile Zhang\orcidlink{0000-0003-1673-5901}}
\email{blzhang@ntu.edu.sg}
\affiliation{Division of Physics and Applied Physics, School of Physical and Mathematical Sciences, Nanyang Technological University,
Singapore 637371, Singapore}
\affiliation{Centre for Disruptive Photonic Technologies, Nanyang Technological University, Singapore 637371, Singapore}

\begin{abstract}
Entanglement entropy scaling links geometry to quantum correlations and diagnoses quantum criticality. Fractal lattices extend this link to noninteger dimensions, where boundary and bulk dimensions are independent. It remains open which dimension governs entropy scaling and how strongly gap closing can enhance it. Nonlocal occupied-state projectors and the lack of translation symmetry obstruct standard methods. We overcome these obstacles and classify free-fermion entropy scaling with bounds set by the hopping across the subsystem boundary: an upper bound from its rank and the spectral gap, and a lower bound from how much it lowers the ground-state energy. With only negative-energy states filled and this energy lowering proportional to the boundary, the entanglement entropy obeys the boundary law $S_X=\Theta\left(L_X^{d_{\mathrm b}}\right)$ under a uniform gap and satisfies $L_X^{d_{\mathrm b}}\lesssim S_X\lesssim L_X^{d_{\mathrm b}}\ln L_X$ when the gap closes no faster than algebraically. Here $L_X$ is the subsystem linear size and $d_{\mathrm b}$ the boundary dimension. Half occupation of the chiral zero modes gives a mixed-state entropy $S_X^{\mathrm{mix}}$ with $\max\left\{S_X,z_X\ln2\right\}\leqslant S_X^{\mathrm{mix}}\leqslant S_X+z_X\ln2$, where $z_X$ is the local zero-mode weight; an extensive $z_X$ thus makes $S_X^{\mathrm{mix}}$ follow the bulk dimension.
\end{abstract}

\maketitle

\mysec{Introduction---}Entanglement entropy (EE) scaling links spatial geometry to quantum correlations and provides a probe of critical behavior~\cite{vidal2003entanglement, eisert2010colloquium}. For example, logarithmic EE scaling~\cite{cheng2026critical} and the entanglement spectrum~\cite{lin2026critical} have been used to diagnose critical topology in experiments. For a subsystem $X$ of linear size $L_X$ in a periodic $d$-dimensional Euclidean lattice, the EE of gapped free fermions obeys the boundary-law bound $S_X=O\left(L_X^{d-1}\right)$, whereas a codimension-one Fermi surface gives $S_X\sim L_X^{d-1}\ln L_X$~\cite{wolf2006violation, gioev2006entanglement, swingle2010fermi}. The rigorous gapless result relies on Widom asymptotics, regular geometry, and translation symmetry~\cite{leschke2014scaling}; localization instead restores a boundary law for disordered free fermions~\cite{elgart2017large}. Beyond ground states, highly excited fermionic eigenstates and generic random states can exhibit volume-law entanglement~\cite{page1993average, vidmar2017quadratic}.

\begin{figure}
\centering
\includegraphics[width=\columnwidth]{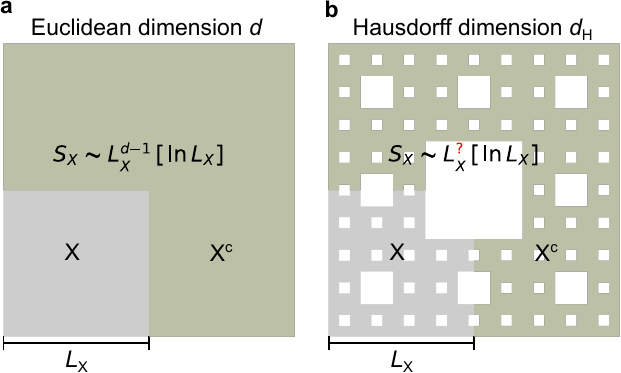}
\caption{\textbf{Entanglement entropy scaling in Euclidean and self-similar lattices.} 
Subsystem $X$ has linear size $L_X$ and complement $X^c$. \textbf{a}, On a periodic $d$-dimensional lattice, gapped free-fermion ground states satisfy $S_X=O\left(L_X^{d-1}\right)$, while a codimension-one Fermi surface gives $S_X\sim L_X^{d-1}\ln L_X$. \textbf{b}, A self-similar lattice with Hausdorff dimension $d_{\mathrm H}$. The red question mark denotes the exponent to be determined; square brackets indicate possible logarithmic enhancement. L1--L3 address this scaling problem through conditions on the cut, spectrum, and occupation.}
\label{Fig_01}
\end{figure}

Fractal lattices have been realized experimentally~\cite{kempkes2019design, xu2021transport, biesenthal2022fractal} and offer a setting for exploring EE. Their self-similar structure can modify conventional boundary-law behavior~\cite{hamma2010fractal, li2022laughlin}, but the resulting scaling cannot be inferred from the fractal dimension alone. The bulk scales as $L_X^{d_{\mathrm H}}$, while an entangling boundary can scale independently as $L_X^{d_{\mathrm b}}$, where $d_{\mathrm H}$ and $d_{\mathrm b}$ are the bulk and boundary dimensions, respectively. In particular, unlike on Euclidean lattices, $d_{\mathrm b}$ need not equal $d_{\mathrm H}-1$ [Fig.~\ref{Fig_01}]. Field theory can treat noninteger dimensions through analytic continuation to spacetime dimension $D=4-\epsilon$, allowing calculations of EE and related sphere free energies~\cite{metlitski2009entanglement, fei2015generalized}. However, this continuation does not incorporate self-similar lattices. Numerical correlation-matrix calculations can directly evaluate EE on finite fractal lattices~\cite{zhou2024entanglement, kosior2026entanglement}, but the rapid growth of the number of sites with generation limits the accessible sizes, so finite-size data cannot settle the asymptotic scaling. It therefore remains open which dimension governs EE scaling on fractal lattices and how strongly gap closing can enhance it [Fig.~\ref{Fig_01}\textbf{b}].

Answering these questions requires bounds on the occupied-state projector restricted to $X$. The full projector can extend across the system even for finite-range hopping. Without translation symmetry, the standard Euclidean asymptotic methods do not directly apply~\cite{gioev2006entanglement, leschke2014scaling}.

In this Letter, we overcome these obstacles and establish a classification of free-fermion entropy scaling on fractal lattices. We bound the entropy using only the hopping across the subsystem boundary and the energy gap. The upper bound scales with the number of independent hopping channels and grows only logarithmically as the gap closes. The lower bound grows with how much this hopping lowers the ground-state energy. Neither bound requires translation symmetry. With only negative-energy states filled and this energy lowering proportional to the boundary, these bounds give $S_X=\Theta\left(L_X^{d_{\mathrm b}}\right)$ for a uniform occupied--unoccupied gap (L1) and $cL_X^{d_{\mathrm b}}\leqslant S_X\leqslant CL_X^{d_{\mathrm b}}\ln L_X$ for at most algebraic gap closure (L2), with size-independent $c,C>0$. Separately, for the grand canonical state at zero temperature and chemical potential, the subsystem entropy $S_X^{\mathrm{mix}}$ obeys the gap-independent bound $\max\left\{S_X,z_X\ln2\right\}\leqslant S_X^{\mathrm{mix}}\leqslant S_X+z_X\ln2$ (L3), where $z_X$ is the local zero-mode weight. These results answer both questions: the boundary dimension governs the pure-state EE, algebraic gap closing adds at most one logarithm, and an extensive zero-mode weight makes the bulk dimension govern $S_X^{\mathrm{mix}}$. We verify L1--L3 on the Sierpi\'nski carpet and Vicsek lattices.

\begin{figure*}
\centering
\includegraphics[width=\textwidth]{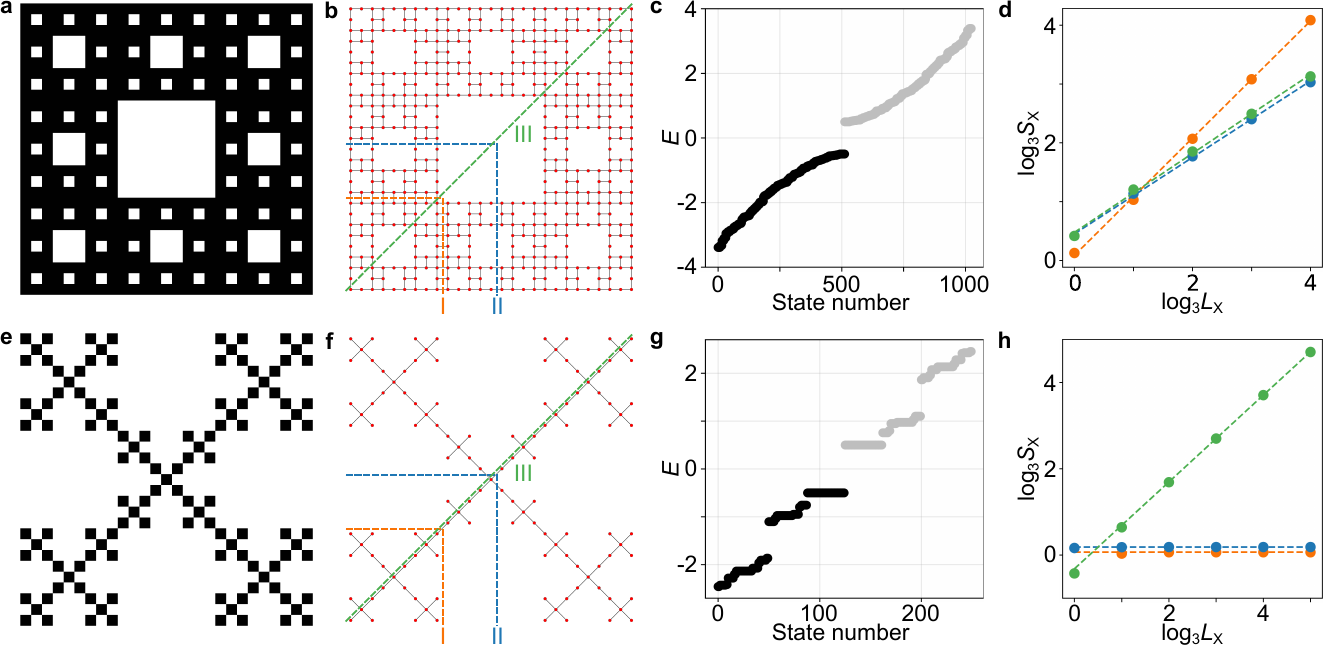}
\caption{\textbf{L1: Negative filling with a uniform nonzero-energy gap.} Panels \textbf{a}--\textbf{c} and \textbf{e}--\textbf{g} show generation $g=3$. \textbf{a,e,} Sierpi\'nski carpet and Vicsek fractal geometries. \textbf{b,f,} Corresponding lattices with cuts I (orange), II (blue), and III (green). \textbf{c,g,} Spectra of Eq.~\eqref{eq_02} with $t=1$ and $t_1=0.5$; occupied (empty) states are black (gray). \textbf{d,h,} EE for the same cuts, with matching colors. Points show $g=1,\ldots,5$ in \textbf{d} and $g=1,\ldots,6$ in \textbf{h}. Dashed lines are power-law fits over $g=1,\ldots,5$ and $g=3,\ldots,6$, respectively.}
\label{Fig_02}
\end{figure*}

\mysec{Entanglement entropy bounds---}For a pure number-conserving Gaussian state, let $P$ be its single-particle correlation projector and $\Pi_X$ the projector onto all local orbitals in subsystem $X$. The boundary between $X$ and its complement $X^c$ defines the cut, with $\Pi_{X^c}=I-\Pi_X$. The eigenvalues $\lambda_\alpha\in[0,1]$ of the restricted correlation matrix $C_X=\Pi_XP\Pi_X$ determine the EE~\cite{peschel2003calculation, peschel2009reduced, zhou2022floquet}
\begin{equation}
S_X=-\sum_\alpha\left[\lambda_\alpha\ln\lambda_\alpha+\left(1-\lambda_\alpha\right)\ln\left(1-\lambda_\alpha\right)\right].
\label{eq_entropy_correlation}
\end{equation}

Let $h$ be a single-particle Hamiltonian with $[h,P]=0$ and $M=\|h\|_{\mathrm{op}}>0$. Thus $M$ is the largest single-particle energy magnitude. The cut hopping matrix $W_X=\Pi_Xh\Pi_{X^c}+\Pi_{X^c}h\Pi_X$ defines the energy gain $\mathcal T_X=-\operatorname{Tr}\left(PW_X\right)$. Let $\mathcal E_{\partial X}$ contain the geometric bonds crossing the cut, each counted once, and let $b_X=\left|\mathcal E_{\partial X}\right|$. For distinct sites, define the density anticorrelation $w_{ij}=\left\langle\hat n_i\right\rangle\left\langle\hat n_j\right\rangle-\left\langle\hat n_i\hat n_j\right\rangle$, where $\hat n_i$ is the total particle-number operator at site $i$ and expectations are taken in the Gaussian state. Then
\begin{equation}
S_X\geqslant(\ln2)\max\left\{
\frac{\mathcal T_X}{M},\,
4\sum_{\{i,j\}\in\mathcal E_{\partial X}}w_{ij}\right\}.
\label{eq_ee_lower_bounds}
\end{equation}
For pure negative filling, $\mathcal T_X$ is the energy cost of removing cut hopping while keeping the state fixed. A boundary lower bound follows if either of the following averages has a positive, size-independent lower bound:
\begin{subequations}
\label{eq_ee_lower_conditions}
\begin{align}
\frac{\mathcal T_X}{b_X}&\geqslant\varepsilon_*>0,
\label{eq_ee_energy_condition}\\
{\frac{1}{b_X}\sum_{\{i,j\}\in\mathcal E_{\partial X}}w_{ij}}&{\geqslant\gamma_*>0.}
\label{eq_ee_correlation_condition}
\end{align}
\end{subequations}
Here $b_X>0$, and $\varepsilon_*$ and $\gamma_*$ are independent of system size; the energy route also requires a size-independent upper bound on $M$. If $b_X=\Theta\left(L_X^{d_{\mathrm b}}\right)$, either condition yields
\begin{equation}
S_X\geqslant cL_X^{d_{\mathrm b}},\qquad c>0,
\label{eq_ee_boundary_lower}
\end{equation}
with $c$ independent of size. Neither lower bound requires a spectral gap.

The upper bound depends on the cut hopping rank $r_X=\operatorname{rank}\left(\Pi_Xh\Pi_{X^c}\right)$ and the spectral separation. If the occupied and unoccupied energies lie in $\left[-M,-\delta_g\right]$ and $[0,M]$, respectively, the block $P\Pi_X(I-P)$ satisfies a Sylvester equation with displacement rank at most $2r_X$. Applying the singular-value estimate of Beckermann and Townsend~\cite{beckermann2017displacement} gives
\begin{equation}
S_X\leqslant8(\ln2)r_X
\left[1+\frac{1}{\pi^2}\ln\left(4+\frac{8M}{\delta_g}\right)\right].
\label{eq_ee_rank_gap_upper}
\end{equation}
Here $\delta_g>0$ bounds the occupied--unoccupied separation from below at generation $g$ and may decrease with $g$. Derivations are given in Secs.~\SM{S1 B}--\SM{S1 E} of~\cite{SM}.

Eqs.~\eqref{eq_ee_lower_conditions}, \eqref{eq_ee_rank_gap_upper} provide the common basis of L1 and L2. For Hamiltonians with uniformly bounded $M$ and $r_X=O\left(b_X\right)$, either condition in Eq.~\eqref{eq_ee_lower_conditions} fixes a positive boundary lower bound; the gap dependence of Eq.~\eqref{eq_ee_rank_gap_upper} then controls the possible enhancement. A uniform gap gives a boundary law, while at most algebraic gap closure permits at most a logarithmic enhancement. L3 changes the zero-mode occupation and uses a separate comparison bound.

\mysec{Models and cuts---}To realize these conditions, we consider two quadratic tight-binding models on finite bipartite graph approximants $V=A\cup B$, with nearest-neighbor bonds joining opposite sublattices. The first is~\cite{zhou2024entanglement}
\begin{equation}
    H=t\sum_{\langle i,j\rangle}
    c_i^\dagger c_j \,,
    \label{eq_01}
\end{equation}
where $c_i^\dagger$ creates a spinless fermion, $t$ is real, and $\langle i,j\rangle$ includes both bond orientations. In the $(A,B)$ basis, the single-particle matrix has the chiral form $h=\left(\begin{smallmatrix}0&T\\T^\dagger&0\end{smallmatrix}\right)$ and anticommutes with $\Gamma=\operatorname{diag}\left(\mathbb{I}_A,-\mathbb{I}_B\right)$. Hence nonzero energies occur in $\pm E$ pairs and $n_0\equiv\dim\ker h\geqslant\left|N_A-N_B\right|$~\cite{inui1994unusual}. For a nonempty nonzero spectrum, define 
\begin{equation}
\delta_g=\min_{E\ne0}|E|=\min_{E<0}|E|.
\end{equation}
We call the nonzero spectrum uniformly gapped when $\inf_g\delta_g>0$ and gapless when $\delta_g\to0$.  Throughout, $S_X$ denotes the EE under pure negative filling, while $S_X^{\mathrm{mix}}$ denotes the mixed-state subsystem entropy under independent half occupation of the zero modes. We set $t=1$.

For an explicitly gapped reference, we use the two-orbital Hamiltonian
\begin{equation}
    H=t_1\sum_i\Psi_i^\dagger\sigma_x\Psi_i
    +t\sum_{\langle i,j\rangle}
    \left(\Psi_i^\dagger\sigma_z\Psi_j \right),
    \qquad
    \Psi_i=\begin{pmatrix}c_{s,i}\\c_{p,i}\end{pmatrix} \,.
    \label{eq_02}
\end{equation}
Here $\Psi_i$ contains the $s$- and $p$-orbital operators, and $\sigma_x$ and $\sigma_z$ act in orbital space. The chiral operator of Eq.~\eqref{eq_02} is $\Gamma=I_V\otimes\sigma_y$, which acts within each physical site. For every eigenvalue $\lambda$ of the graph adjacency matrix, $E_{\lambda,\pm}=\pm\sqrt{t_1^2+t^2\lambda^2}$; therefore $t_1\neq0$ guarantees a graph-independent half-filling gap of at least $2\left|t_1\right|$. The ground state fills the negative-energy band. We take $t=1$ and $t_1=0.5$.

We test both Hamiltonians on graph approximants of the Sierpi\'nski carpet and Vicsek fractal. At every iteration, each retained square is divided into a $3\times3$ array. The Sierpi\'nski carpet retains all eight subsquares except the center [Fig.~\ref{Fig_02}\textbf{a}], whereas the Vicsek fractal construction retains the center and four corners [Fig.~\ref{Fig_02}\textbf{e}]. At generation $g$, we place one lattice site at the center of every smallest retained square and join neighboring centers---edge-sharing cells for the Sierpi\'nski carpet and corner-sharing cells for the Vicsek fractal---to obtain the graphs in Figs.~\ref{Fig_02}\textbf{b} and \ref{Fig_02}\textbf{f}. Taking the smallest-cell side as the unit length, the overall side length is $L=3^g$, and the total site number $N_g$ is $8^g$ for the Sierpi\'nski carpet and $5^g$ for the Vicsek lattice. Their Hausdorff dimensions are therefore $d_{\mathrm H}=\log_3 8$ and $d_{\mathrm H}=\log_3 5$, respectively.

With the lower-left corner as the origin and $(x,y)$ the site coordinates, define three subsystems: $X_{\mathrm I}=\left\{(x,y):x\leqslant L/3,\ y\leqslant L/3\right\}$, $X_{\mathrm{II}}=\left\{(x,y):x\leqslant L/2,\ y\leqslant L/2\right\}$, and $X_{\mathrm{III}}=\left\{(x,y):y\leqslant x\right\}$. Their boundaries are labeled I--III in Figs.~\ref{Fig_02}\textbf{b} and \ref{Fig_02}\textbf{f}. All three linear sizes are proportional to $L$; we use $L_X=L/3=3^{g-1}$, since fixed rescalings leave the exponents unchanged.

\mysec{L1: Negative-filling entropy with a uniform nonzero-energy gap---}
Consider any Hamiltonian with uniformly bounded $M$ and $r_X=O\left(b_X\right)$. A uniform gap $\delta_g\geqslant\delta_*>0$ turns Eq.~\eqref{eq_ee_rank_gap_upper} into $S_X=O\left(b_X\right)$. Either lower-bound condition in Eq.~\eqref{eq_ee_lower_conditions} then gives the boundary law $S_X=\Theta\left(b_X\right)$. When $b_X=\Theta\left(L_X^{d_{\mathrm b}}\right)$, it follows that $S_X=\Theta\left(L_X^{d_{\mathrm b}}\right)$.

Under these conditions, for finite-range Hamiltonians on bounded-degree graphs with a fixed number of orbitals per site, convergence of the boundary content and local boundary-pattern frequencies further gives
\begin{equation}
    S_X=C_\partial L_X^{d_{\mathrm b}}
    +o\left(L_X^{d_{\mathrm b}}\right),\qquad C_\partial>0.
    \label{eq_r1_law}
\end{equation}
The coefficient conditions and proof are given in Secs.~\SM{S1 G}--\SM{S1 H} of~\cite{SM}.

For a concrete realization, the two-orbital model in Eq.~\eqref{eq_02} supplies both the gap and the positive lower bound: for fixed $t,t_1\ne0$, $\delta_g\geqslant\left|t_1\right|$ and $\mathcal T_X/b_X\geqslant2t^2t_1^2/\left(t_1^2+t^2d_*^2\right)^{3/2}>0$, where $d_*$ is a generation-independent upper bound on the vertex degree. It therefore obeys the boundary law, providing an analytical basis for the gapped Sierpi\'nski carpet scaling previously inferred numerically~\cite{zhou2024entanglement}. The model verification is given in Sec.~\SM{S1 F} of~\cite{SM}.

Figure~\ref{Fig_02}\textbf{d} gives fitted exponents $(0.9970,0.6501,0.6715)$ for Sierpi\'nski carpet cuts I--III, consistent with $\left(1,\log_3 2,\log_3 2\right)$, where $\log_3 2\approx 0.6309$. The Vicsek fractal fits are $(0.000039,0.000040,1.0065)$, consistent with $(0,0,1)$ [Fig.~\ref{Fig_02}\textbf{h}]: the two bottleneck cuts approach saturation, while the third grows with its boundary. These results show how boundary geometry selects the EE exponent for the same gapped Hamiltonian. Further model tests and fit diagnostics are given in Secs.~\SM{S1 I}--\SM{S1 K} of~\cite{SM}.

\mysec{L2: Negative-filling entropy under gap closure---}
Retain the uniform bounds on $M$ and $r_X/b_X$, but let $\delta_g\to0$ with $\delta_g\geqslant c_\delta L_X^{-\zeta}$ for fixed $c_\delta>0$ and $0<\zeta<\infty$. This condition includes the algebraic gap scaling familiar from finite-size scaling at continuous quantum transitions~\cite{campostrini2014finite}. Combining Eq.~\eqref{eq_ee_rank_gap_upper} with either condition in Eq.~\eqref{eq_ee_lower_conditions} yields
\begin{equation}
 cb_X\leqslant S_X\leqslant Cb_X\ln L_X,
 \label{eq_l2_general_bounds}
\end{equation}
for sufficiently large generations, with size-independent $c,C>0$. These model-independent bounds constrain the allowed scaling but leave the leading logarithmic factor undetermined; the derivation is given in Sec.~\SM{S2 A} of~\cite{SM}.

The single-orbital model in Eq.~\eqref{eq_01} on the uniform Sierpi\'nski carpet provides a sharper application. We prove $\delta_g=\Theta\left(L_X^{-1}\right)$ and both conditions in Eq.~\eqref{eq_ee_lower_conditions} for cuts I--III. Combining exact dimer elimination with local correlation estimates and convergent boundary-pattern frequencies establishes the asymptotic boundary law
\begin{equation}
 S_X =  C_{\partial,X}L_X^{d_{\mathrm b}} + o\left(L_X^{d_{\mathrm b}}\right) ,\qquad 0<C_{\partial,X}<\infty.
 \label{eq_r3_law}
\end{equation}
Here $C_{\partial,X}$ is independent of generation and depends on the cut; the boundary exponents are $\left(1,\log_3 2,\log_3 2\right)$. Thus the gapless carpet realizes a boundary law within the general L2 bounds, with a definite leading coefficient and no multiplicative logarithmic enhancement. Dimer elimination maps the carpet onto a hierarchy of even cycles. Each cycle contributes at most a logarithm of its length, but the number of long cycles crossing the cut per boundary bond decreases as a power of their length, so the total remains proportional to the boundary. Proofs are given in Secs.~\SM{S2 B}--\SM{S2 D} of~\cite{SM}.

\begin{figure}
\centering
\includegraphics[width=\columnwidth]{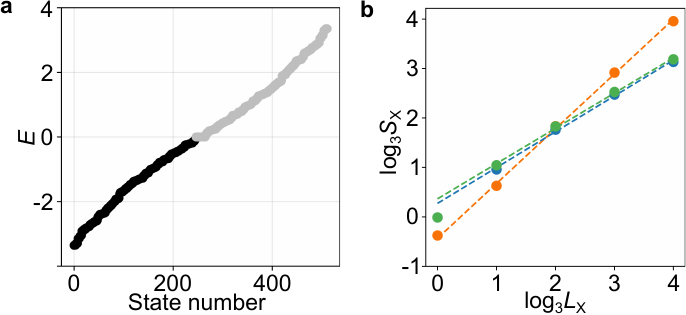}
\caption{\textbf{L2: Negative filling on the gapless Sierpi\'nski carpet.} \textbf{a,} Spectrum of Eq.~\eqref{eq_01} at $g=3$. Negative-energy states (black) are filled; zero- and positive-energy states (gray) are empty. \textbf{b,} EE $S_X$ for cuts I--III of Fig.~\ref{Fig_02}\textbf{b}, with matching colors. Points show $g=1,\ldots,5$; dashed lines are power-law fits over $g=2,\ldots,5$.}
\label{Fig_03}
\end{figure}

Figure~\ref{Fig_03}\textbf{a} shows the pure negative-filling spectrum. Fits to the EE in Fig.~\ref{Fig_03}\textbf{b} over $g=2,\ldots,5$ give finite-size effective exponents $(1.1071,0.7241,0.7126)$ for cuts I--III. The excess over $d_{\mathrm b}$ comes from a negative subleading correction to the boundary law, whose relative size falls by about half per generation. The asymptotic boundary exponents and the absence of a leading logarithm follow from the analytical argument above. Detailed finite-size and fitting-window analyses are given in Sec.~\SM{S2 E} of~\cite{SM}.

\begin{figure}
\centering
\includegraphics[width=\columnwidth]{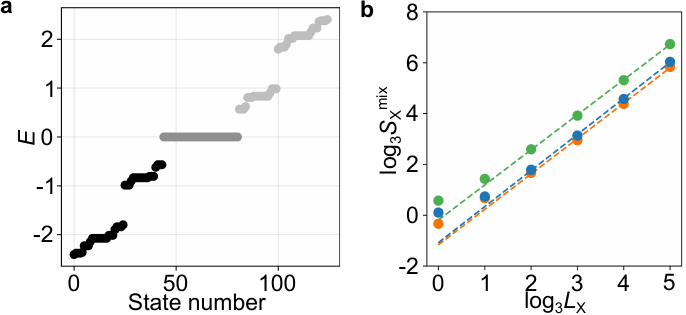}
\caption{\textbf{L3: Mixed half filling on the Vicsek lattice.} \textbf{a,} Spectrum of Eq.~\eqref{eq_01} at $g=3$. Negative-energy states (black) are filled, zero modes (dark gray) are independently occupied with probability $1/2$, and positive-energy states (light gray) are empty. \textbf{b,} Mixed-state subsystem entropy $S_X^{\mathrm{mix}}$ for cuts I--III of Fig.~\ref{Fig_02}\textbf{f}, with matching colors. Points show $g=1,\ldots,6$; dashed lines are power-law fits over $g=3,\ldots,6$.}
\label{Fig_04}
\end{figure}

\mysec{L3: Mixed-state entropy from half-occupied zero modes---}
The third part of the classification changes the occupation of exact zero modes. At fixed generation, the zero-temperature grand canonical state at zero chemical potential independently occupies each exact zero mode with probability $1/2$, forming a statistical mixture of many-body states with different particle numbers. With $P_-=\mathbf1(h<0)$ and $Z=\mathbf1(h=0)$, its correlation operator is $P_{\rm mix}=P_-+\frac12Z$. Define the local zero-mode weight $z_X=\operatorname{Tr}\left(\Pi_XZ\Pi_X\right)$. For any Hamiltonian with site-local chiral symmetry, the pure negative-filling EE $S_X$ and mixed-state subsystem entropy $S_X^{\mathrm{mix}}$ satisfy
\begin{equation}
 \max\left\{S_X,z_X\ln2\right\}\leqslant S_X^{\mathrm{mix}}
 \leqslant S_X+z_X\ln2.
 \label{eq_r2_weight_bound}
\end{equation}
This bound holds at every finite generation, independently of the spectral gap. It classifies the mixed entropy according to the relative growth of $z_X$ and $S_X$:
\begin{equation}
 S_X^{\mathrm{mix}}=
 \begin{cases}
 S_X+o\left(S_X\right),&z_X=o\left(S_X\right),\\
 \Theta\left(S_X\right),&z_X=\Theta\left(S_X\right),\\
 z_X\ln2+o\left(z_X\right),&S_X=o\left(z_X\right).
 \end{cases}
 \label{eq_l3_competition}
\end{equation}
The term $z_X\ln2$ is the local weight of thermodynamic occupation entropy; the derivation is given in Sec.~\SM{S3 A} of~\cite{SM}.

When $S_X=o\left(L_X^{d_{\mathrm H}}\right)$ and $z_X/L_X^{d_{\mathrm H}}\to a_X>0$, the last regime yields
\begin{equation}
 S_X^{\mathrm{mix}}=C_{\rm mix}L_X^{d_{\mathrm H}}+o\left(L_X^{d_{\mathrm H}}\right),
 \label{eq_l3_coefficient}
\end{equation}
where $C_{\rm mix}=a_X\ln2>0$ is independent of generation but may depend on the cut [Sec.~\SM{S3 D} of~\cite{SM}].

For the nearest-neighbor Vicsek fractal in Fig.~\ref{Fig_04}, recursive counting gives $z_X/|X|\to7/25$ and $|X|\sim A_XL_X^{d_{\mathrm H}}$, with $A_X=(1,5/4,5/2)$ for cuts I--III. A tree spectral bound gives $S_X=O\left(b_X\ln L_X\right)=o(|X|)$, verifying both conditions and fixing $C_{\rm mix}=7A_X\ln2/25$. Fits over $g=3,\ldots,6$ give finite-size effective exponents $(1.3926,1.4190,1.3805)$ [Fig.~\ref{Fig_04}\textbf{b}], below the analytically established bulk exponent $d_{\mathrm H}=\log_3 5\simeq1.4650$. Proofs, fit diagnostics, and further model tests are given in Secs.~\SM{S3 E}--\SM{S3 G} of~\cite{SM}.

\mysec{Conclusion---}Our classification provides an analytical framework for comparing entropy scaling across fractal lattices, cuts, and particle occupations. By relating nonlocal entropy to boundary couplings, spectral gaps, and local zero-mode weight, we derive scaling laws across free-fermion Hamiltonians without relying on translation symmetry. This approach establishes infinite-size behavior beyond numerically accessible generations, providing a physical basis for interpreting finite-size entropy data.

Future work should identify the conditions under which gap closure produces a leading logarithmic enhancement of entanglement entropy. For mixed states, a further goal is to quantify quantum correlations independently of thermodynamic occupation entropy. Nonbipartite models without chiral symmetry offer a direct test of the role of symmetry in entropy scaling~\cite{brzezinska2018topology}. Random fractals provide a setting to study how localization competes with cut geometry~\cite{kosior2026entanglement}, while multifractal geometries raise the question of whether a single boundary exponent suffices to describe the entropy. In three-dimensional fractals, localization transitions open further questions about how the spatial structure of critical eigenstates shapes EE and whether it produces distinct scaling laws~\cite{li2026anderson}.
 
\mysec{Acknowledgments---}This work is supported by Singapore National Research Foundation Thematic Competitive Research Program under Grant No. NRF-T-CRP-2025-0004, and Singapore Ministry of Education Academic Research Fund Tier 2 under Grant No. MOE-T2EP50123-0007, and Tier 1 Grant RG176/25. 
Baile Zhang acknowledges support from Nanyang Technological University. 
Yihao Yang is supported by Key Research and Development Program of the Ministry of Science and Technology under Grant No. 2022YFA1404900, 2022YFA1405200, National Natural Science Foundation of China (NSFC) under Grants U25D8017, and Class D for Young Scientific Research Peak Creation No. K20250230.

We used Qiushi Discovery Engine~\cite{yang2026qiushi} and OpenAI Codex to assist mathematical derivations.

\mysec{Data availability---}The numerical codes and data in this study have been deposited in the Nanyang Technological University database under accession code [insert upon publication].

\twocolumngrid
\mysec{References}
\makeatletter
\renewcommand{\bibsection}{}
\makeatother


\begin{thebibliography}{40}%
\makeatletter
\providecommand \@ifxundefined [1]{%
 \@ifx{#1\undefined}
}%
\providecommand \@ifnum [1]{%
 \ifnum #1\expandafter \@firstoftwo
 \else \expandafter \@secondoftwo
 \fi
}%
\providecommand \@ifx [1]{%
 \ifx #1\expandafter \@firstoftwo
 \else \expandafter \@secondoftwo
 \fi
}%
\providecommand \natexlab [1]{#1}%
\providecommand \enquote  [1]{``#1''}%
\providecommand \bibnamefont  [1]{#1}%
\providecommand \bibfnamefont [1]{#1}%
\providecommand \citenamefont [1]{#1}%
\providecommand \href@noop [0]{\@secondoftwo}%
\providecommand \href [0]{\begingroup \@sanitize@url \@href}%
\providecommand \@href[1]{\@@startlink{#1}\@@href}%
\providecommand \@@href[1]{\endgroup#1\@@endlink}%
\providecommand \@sanitize@url [0]{\catcode `\\12\catcode `\$12\catcode
  `\&12\catcode `\#12\catcode `\^12\catcode `\_12\catcode `\%12\relax}%
\providecommand \@@startlink[1]{}%
\providecommand \@@endlink[0]{}%
\providecommand \url  [0]{\begingroup\@sanitize@url \@url }%
\providecommand \@url [1]{\endgroup\@href {#1}{\urlprefix }}%
\providecommand \urlprefix  [0]{URL }%
\providecommand \Eprint [0]{\href }%
\providecommand \doibase [0]{https://doi.org/}%
\providecommand \selectlanguage [0]{\@gobble}%
\providecommand \bibinfo  [0]{\@secondoftwo}%
\providecommand \bibfield  [0]{\@secondoftwo}%
\providecommand \translation [1]{[#1]}%
\providecommand \BibitemOpen [0]{}%
\providecommand \bibitemStop [0]{}%
\providecommand \bibitemNoStop [0]{.\EOS\space}%
\providecommand \EOS [0]{\spacefactor3000\relax}%
\providecommand \BibitemShut  [1]{\csname bibitem#1\endcsname}%
\let\auto@bib@innerbib\@empty
\bibitem [{\citenamefont {Vidal}\ \emph {et~al.}(2003)\citenamefont {Vidal},
  \citenamefont {Latorre}, \citenamefont {Rico},\ and\ \citenamefont
  {Kitaev}}]{vidal2003entanglement}%
  \BibitemOpen
  \bibfield  {author} {\bibinfo {author} {\bibfnamefont {G.}~\bibnamefont
  {Vidal}}, \bibinfo {author} {\bibfnamefont {J.~I.}\ \bibnamefont {Latorre}},
  \bibinfo {author} {\bibfnamefont {E.}~\bibnamefont {Rico}},\ and\ \bibinfo
  {author} {\bibfnamefont {A.}~\bibnamefont {Kitaev}},\ }\bibfield  {title}
  {\bibinfo {title} {Entanglement in quantum critical phenomena},\ }\href
  {https://doi.org/10.1103/PhysRevLett.90.227902} {\bibfield  {journal}
  {\bibinfo  {journal} {Phys. Rev. Lett.}\ }\textbf {\bibinfo {volume} {90}},\
  \bibinfo {pages} {227902} (\bibinfo {year} {2003})}\BibitemShut {NoStop}%
\bibitem [{\citenamefont {Eisert}\ \emph {et~al.}(2010)\citenamefont {Eisert},
  \citenamefont {Cramer},\ and\ \citenamefont {Plenio}}]{eisert2010colloquium}%
  \BibitemOpen
  \bibfield  {author} {\bibinfo {author} {\bibfnamefont {J.}~\bibnamefont
  {Eisert}}, \bibinfo {author} {\bibfnamefont {M.}~\bibnamefont {Cramer}},\
  and\ \bibinfo {author} {\bibfnamefont {M.~B.}\ \bibnamefont {Plenio}},\
  }\bibfield  {title} {\bibinfo {title} {Colloquium: Area laws for the
  entanglement entropy},\ }\href {https://doi.org/10.1103/RevModPhys.82.277}
  {\bibfield  {journal} {\bibinfo  {journal} {Rev. Mod. Phys.}\ }\textbf
  {\bibinfo {volume} {82}},\ \bibinfo {pages} {277} (\bibinfo {year}
  {2010})}\BibitemShut {NoStop}%
\bibitem [{\citenamefont {Cheng}\ \emph {et~al.}(2026)\citenamefont {Cheng},
  \citenamefont {Zhang}, \citenamefont {Yu}, \citenamefont {Zhou},
  \citenamefont {Gong},\ and\ \citenamefont {Zhang}}]{cheng2026critical}%
  \BibitemOpen
  \bibfield  {author} {\bibinfo {author} {\bibfnamefont {Z.}~\bibnamefont
  {Cheng}}, \bibinfo {author} {\bibfnamefont {X.}~\bibnamefont {Zhang}},
  \bibinfo {author} {\bibfnamefont {X.-J.}\ \bibnamefont {Yu}}, \bibinfo
  {author} {\bibfnamefont {L.}~\bibnamefont {Zhou}}, \bibinfo {author}
  {\bibfnamefont {J.}~\bibnamefont {Gong}},\ and\ \bibinfo {author}
  {\bibfnamefont {B.}~\bibnamefont {Zhang}},\ }\bibfield  {title} {\bibinfo
  {title} {Observation of critical topological phase transition},\ }\href
  {https://doi.org/10.1038/s41586-026-11067-5} {\bibfield  {journal} {\bibinfo
  {journal} {Nature}\ }\textbf {\bibinfo {volume} {658}},\ \bibinfo {pages}
  {358} (\bibinfo {year} {2026})}\BibitemShut {NoStop}%
\bibitem [{\citenamefont {Lin}\ \emph {et~al.}(2026)\citenamefont {Lin},
  \citenamefont {Wang}, \citenamefont {Kong}, \citenamefont {Zhou},
  \citenamefont {Lin}, \citenamefont {Yu}, \citenamefont {Zhang},\ and\
  \citenamefont {Jiang}}]{lin2026critical}%
  \BibitemOpen
  \bibfield  {author} {\bibinfo {author} {\bibfnamefont {Z.-K.}\ \bibnamefont
  {Lin}}, \bibinfo {author} {\bibfnamefont {L.-W.}\ \bibnamefont {Wang}},
  \bibinfo {author} {\bibfnamefont {Z.-L.}\ \bibnamefont {Kong}}, \bibinfo
  {author} {\bibfnamefont {Y.}~\bibnamefont {Zhou}}, \bibinfo {author}
  {\bibfnamefont {H.-Q.}\ \bibnamefont {Lin}}, \bibinfo {author} {\bibfnamefont
  {X.}~\bibnamefont {Yu}}, \bibinfo {author} {\bibfnamefont {S.}~\bibnamefont
  {Zhang}},\ and\ \bibinfo {author} {\bibfnamefont {J.-H.}\ \bibnamefont
  {Jiang}},\ }\bibfield  {title} {\bibinfo {title} {Experimental observation of
  critical topology},\ }\href {https://doi.org/10.1038/s41586-026-11099-x}
  {\bibfield  {journal} {\bibinfo  {journal} {Nature}\ }\textbf {\bibinfo
  {volume} {658}},\ \bibinfo {pages} {365} (\bibinfo {year}
  {2026})}\BibitemShut {NoStop}%
\bibitem [{\citenamefont {Wolf}(2006)}]{wolf2006violation}%
  \BibitemOpen
  \bibfield  {author} {\bibinfo {author} {\bibfnamefont {M.~M.}\ \bibnamefont
  {Wolf}},\ }\bibfield  {title} {\bibinfo {title} {Violation of the entropic
  area law for fermions},\ }\href
  {https://doi.org/10.1103/PhysRevLett.96.010404} {\bibfield  {journal}
  {\bibinfo  {journal} {Phys. Rev. Lett.}\ }\textbf {\bibinfo {volume} {96}},\
  \bibinfo {pages} {010404} (\bibinfo {year} {2006})}\BibitemShut {NoStop}%
\bibitem [{\citenamefont {Gioev}\ and\ \citenamefont
  {Klich}(2006)}]{gioev2006entanglement}%
  \BibitemOpen
  \bibfield  {author} {\bibinfo {author} {\bibfnamefont {D.}~\bibnamefont
  {Gioev}}\ and\ \bibinfo {author} {\bibfnamefont {I.}~\bibnamefont {Klich}},\
  }\bibfield  {title} {\bibinfo {title} {Entanglement entropy of fermions in
  any dimension and the {Widom} conjecture},\ }\href
  {https://doi.org/10.1103/PhysRevLett.96.100503} {\bibfield  {journal}
  {\bibinfo  {journal} {Phys. Rev. Lett.}\ }\textbf {\bibinfo {volume} {96}},\
  \bibinfo {pages} {100503} (\bibinfo {year} {2006})}\BibitemShut {NoStop}%
\bibitem [{\citenamefont {Swingle}(2010)}]{swingle2010fermi}%
  \BibitemOpen
  \bibfield  {author} {\bibinfo {author} {\bibfnamefont {B.}~\bibnamefont
  {Swingle}},\ }\bibfield  {title} {\bibinfo {title} {Entanglement entropy and
  the {Fermi} surface},\ }\href
  {https://doi.org/10.1103/PhysRevLett.105.050502} {\bibfield  {journal}
  {\bibinfo  {journal} {Phys. Rev. Lett.}\ }\textbf {\bibinfo {volume} {105}},\
  \bibinfo {pages} {050502} (\bibinfo {year} {2010})}\BibitemShut {NoStop}%
\bibitem [{\citenamefont {Leschke}\ \emph {et~al.}(2014)\citenamefont
  {Leschke}, \citenamefont {Sobolev},\ and\ \citenamefont
  {Spitzer}}]{leschke2014scaling}%
  \BibitemOpen
  \bibfield  {author} {\bibinfo {author} {\bibfnamefont {H.}~\bibnamefont
  {Leschke}}, \bibinfo {author} {\bibfnamefont {A.~V.}\ \bibnamefont
  {Sobolev}},\ and\ \bibinfo {author} {\bibfnamefont {W.}~\bibnamefont
  {Spitzer}},\ }\bibfield  {title} {\bibinfo {title} {Scaling of {R}{\'e}nyi
  entanglement entropies of the free {Fermi}-gas ground state: A rigorous
  proof},\ }\href {https://doi.org/10.1103/PhysRevLett.112.160403} {\bibfield
  {journal} {\bibinfo  {journal} {Phys. Rev. Lett.}\ }\textbf {\bibinfo
  {volume} {112}},\ \bibinfo {pages} {160403} (\bibinfo {year}
  {2014})}\BibitemShut {NoStop}%
\bibitem [{\citenamefont {Elgart}\ \emph {et~al.}(2017)\citenamefont {Elgart},
  \citenamefont {Pastur},\ and\ \citenamefont {Shcherbina}}]{elgart2017large}%
  \BibitemOpen
  \bibfield  {author} {\bibinfo {author} {\bibfnamefont {A.}~\bibnamefont
  {Elgart}}, \bibinfo {author} {\bibfnamefont {L.}~\bibnamefont {Pastur}},\
  and\ \bibinfo {author} {\bibfnamefont {M.}~\bibnamefont {Shcherbina}},\
  }\bibfield  {title} {\bibinfo {title} {Large block properties of the
  entanglement entropy of free disordered fermions},\ }\href
  {https://doi.org/10.1007/s10955-016-1656-z} {\bibfield  {journal} {\bibinfo
  {journal} {J. Stat. Phys.}\ }\textbf {\bibinfo {volume} {166}},\ \bibinfo
  {pages} {1092} (\bibinfo {year} {2017})}\BibitemShut {NoStop}%
\bibitem [{\citenamefont {Page}(1993)}]{page1993average}%
  \BibitemOpen
  \bibfield  {author} {\bibinfo {author} {\bibfnamefont {D.~N.}\ \bibnamefont
  {Page}},\ }\bibfield  {title} {\bibinfo {title} {Average entropy of a
  subsystem},\ }\href {https://doi.org/10.1103/PhysRevLett.71.1291} {\bibfield
  {journal} {\bibinfo  {journal} {Phys. Rev. Lett.}\ }\textbf {\bibinfo
  {volume} {71}},\ \bibinfo {pages} {1291} (\bibinfo {year}
  {1993})}\BibitemShut {NoStop}%
\bibitem [{\citenamefont {Vidmar}\ \emph {et~al.}(2017)\citenamefont {Vidmar},
  \citenamefont {Hackl}, \citenamefont {Bianchi},\ and\ \citenamefont
  {Rigol}}]{vidmar2017quadratic}%
  \BibitemOpen
  \bibfield  {author} {\bibinfo {author} {\bibfnamefont {L.}~\bibnamefont
  {Vidmar}}, \bibinfo {author} {\bibfnamefont {L.}~\bibnamefont {Hackl}},
  \bibinfo {author} {\bibfnamefont {E.}~\bibnamefont {Bianchi}},\ and\ \bibinfo
  {author} {\bibfnamefont {M.}~\bibnamefont {Rigol}},\ }\bibfield  {title}
  {\bibinfo {title} {Entanglement entropy of eigenstates of quadratic fermionic
  hamiltonians},\ }\href {https://doi.org/10.1103/PhysRevLett.119.020601}
  {\bibfield  {journal} {\bibinfo  {journal} {Phys. Rev. Lett.}\ }\textbf
  {\bibinfo {volume} {119}},\ \bibinfo {pages} {020601} (\bibinfo {year}
  {2017})}\BibitemShut {NoStop}%
\bibitem [{\citenamefont {Kempkes}\ \emph {et~al.}(2019)\citenamefont
  {Kempkes}, \citenamefont {Slot}, \citenamefont {Freeney}, \citenamefont
  {Zevenhuizen}, \citenamefont {Vanmaekelbergh}, \citenamefont {Swart},\ and\
  \citenamefont {Morais~Smith}}]{kempkes2019design}%
  \BibitemOpen
  \bibfield  {author} {\bibinfo {author} {\bibfnamefont {S.~N.}\ \bibnamefont
  {Kempkes}}, \bibinfo {author} {\bibfnamefont {M.~R.}\ \bibnamefont {Slot}},
  \bibinfo {author} {\bibfnamefont {S.~E.}\ \bibnamefont {Freeney}}, \bibinfo
  {author} {\bibfnamefont {S.~J.~M.}\ \bibnamefont {Zevenhuizen}}, \bibinfo
  {author} {\bibfnamefont {D.}~\bibnamefont {Vanmaekelbergh}}, \bibinfo
  {author} {\bibfnamefont {I.}~\bibnamefont {Swart}},\ and\ \bibinfo {author}
  {\bibfnamefont {C.}~\bibnamefont {Morais~Smith}},\ }\bibfield  {title}
  {\bibinfo {title} {Design and characterization of electrons in a fractal
  geometry},\ }\href {https://doi.org/10.1038/s41567-018-0328-0} {\bibfield
  {journal} {\bibinfo  {journal} {Nat. Phys.}\ }\textbf {\bibinfo {volume}
  {15}},\ \bibinfo {pages} {127} (\bibinfo {year} {2019})}\BibitemShut
  {NoStop}%
\bibitem [{\citenamefont {Xu}\ \emph {et~al.}(2021)\citenamefont {Xu},
  \citenamefont {Wang}, \citenamefont {Chen}, \citenamefont {Morais~Smith},\
  and\ \citenamefont {Jin}}]{xu2021transport}%
  \BibitemOpen
  \bibfield  {author} {\bibinfo {author} {\bibfnamefont {X.-Y.}\ \bibnamefont
  {Xu}}, \bibinfo {author} {\bibfnamefont {X.-W.}\ \bibnamefont {Wang}},
  \bibinfo {author} {\bibfnamefont {D.-Y.}\ \bibnamefont {Chen}}, \bibinfo
  {author} {\bibfnamefont {C.}~\bibnamefont {Morais~Smith}},\ and\ \bibinfo
  {author} {\bibfnamefont {X.-M.}\ \bibnamefont {Jin}},\ }\bibfield  {title}
  {\bibinfo {title} {Quantum transport in fractal networks},\ }\href
  {https://doi.org/10.1038/s41566-021-00845-4} {\bibfield  {journal} {\bibinfo
  {journal} {Nat. Photonics}\ }\textbf {\bibinfo {volume} {15}},\ \bibinfo
  {pages} {703} (\bibinfo {year} {2021})}\BibitemShut {NoStop}%
\bibitem [{\citenamefont {Biesenthal}\ \emph {et~al.}(2022)\citenamefont
  {Biesenthal}, \citenamefont {Maczewsky}, \citenamefont {Yang}, \citenamefont
  {Kremer}, \citenamefont {Segev}, \citenamefont {Szameit},\ and\ \citenamefont
  {Heinrich}}]{biesenthal2022fractal}%
  \BibitemOpen
  \bibfield  {author} {\bibinfo {author} {\bibfnamefont {T.}~\bibnamefont
  {Biesenthal}}, \bibinfo {author} {\bibfnamefont {L.~J.}\ \bibnamefont
  {Maczewsky}}, \bibinfo {author} {\bibfnamefont {Z.}~\bibnamefont {Yang}},
  \bibinfo {author} {\bibfnamefont {M.}~\bibnamefont {Kremer}}, \bibinfo
  {author} {\bibfnamefont {M.}~\bibnamefont {Segev}}, \bibinfo {author}
  {\bibfnamefont {A.}~\bibnamefont {Szameit}},\ and\ \bibinfo {author}
  {\bibfnamefont {M.}~\bibnamefont {Heinrich}},\ }\bibfield  {title} {\bibinfo
  {title} {Fractal photonic topological insulators},\ }\href
  {https://doi.org/10.1126/science.abm2842} {\bibfield  {journal} {\bibinfo
  {journal} {Science}\ }\textbf {\bibinfo {volume} {376}},\ \bibinfo {pages}
  {1114} (\bibinfo {year} {2022})}\BibitemShut {NoStop}%
\bibitem [{\citenamefont {Hamma}\ \emph {et~al.}(2010)\citenamefont {Hamma},
  \citenamefont {Lidar},\ and\ \citenamefont {Severini}}]{hamma2010fractal}%
  \BibitemOpen
  \bibfield  {author} {\bibinfo {author} {\bibfnamefont {A.}~\bibnamefont
  {Hamma}}, \bibinfo {author} {\bibfnamefont {D.~A.}\ \bibnamefont {Lidar}},\
  and\ \bibinfo {author} {\bibfnamefont {S.}~\bibnamefont {Severini}},\
  }\bibfield  {title} {\bibinfo {title} {Entanglement and area law with a
  fractal boundary in a topologically ordered phase},\ }\href
  {https://doi.org/10.1103/PhysRevA.81.010102} {\bibfield  {journal} {\bibinfo
  {journal} {Phys. Rev. A}\ }\textbf {\bibinfo {volume} {81}},\ \bibinfo
  {pages} {010102(R)} (\bibinfo {year} {2010})}\BibitemShut {NoStop}%
\bibitem [{\citenamefont {Li}\ \emph {et~al.}(2022)\citenamefont {Li},
  \citenamefont {Jha},\ and\ \citenamefont {Nielsen}}]{li2022laughlin}%
  \BibitemOpen
  \bibfield  {author} {\bibinfo {author} {\bibfnamefont {X.}~\bibnamefont
  {Li}}, \bibinfo {author} {\bibfnamefont {M.~C.}\ \bibnamefont {Jha}},\ and\
  \bibinfo {author} {\bibfnamefont {A.~E.~B.}\ \bibnamefont {Nielsen}},\
  }\bibfield  {title} {\bibinfo {title} {Laughlin topology on fractal lattices
  without area-law entanglement},\ }\href
  {https://doi.org/10.1103/PhysRevB.105.085152} {\bibfield  {journal} {\bibinfo
   {journal} {Phys. Rev. B}\ }\textbf {\bibinfo {volume} {105}},\ \bibinfo
  {pages} {085152} (\bibinfo {year} {2022})}\BibitemShut {NoStop}%
\bibitem [{\citenamefont {Metlitski}\ \emph {et~al.}(2009)\citenamefont
  {Metlitski}, \citenamefont {Fuertes},\ and\ \citenamefont
  {Sachdev}}]{metlitski2009entanglement}%
  \BibitemOpen
  \bibfield  {author} {\bibinfo {author} {\bibfnamefont {M.~A.}\ \bibnamefont
  {Metlitski}}, \bibinfo {author} {\bibfnamefont {C.~A.}\ \bibnamefont
  {Fuertes}},\ and\ \bibinfo {author} {\bibfnamefont {S.}~\bibnamefont
  {Sachdev}},\ }\bibfield  {title} {\bibinfo {title} {Entanglement entropy in
  the {$O(N)$} model},\ }\href {https://doi.org/10.1103/PhysRevB.80.115122}
  {\bibfield  {journal} {\bibinfo  {journal} {Phys. Rev. B}\ }\textbf {\bibinfo
  {volume} {80}},\ \bibinfo {pages} {115122} (\bibinfo {year}
  {2009})}\BibitemShut {NoStop}%
\bibitem [{\citenamefont {Fei}\ \emph {et~al.}(2015)\citenamefont {Fei},
  \citenamefont {Giombi}, \citenamefont {Klebanov},\ and\ \citenamefont
  {Tarnopolsky}}]{fei2015generalized}%
  \BibitemOpen
  \bibfield  {author} {\bibinfo {author} {\bibfnamefont {L.}~\bibnamefont
  {Fei}}, \bibinfo {author} {\bibfnamefont {S.}~\bibnamefont {Giombi}},
  \bibinfo {author} {\bibfnamefont {I.~R.}\ \bibnamefont {Klebanov}},\ and\
  \bibinfo {author} {\bibfnamefont {G.}~\bibnamefont {Tarnopolsky}},\
  }\bibfield  {title} {\bibinfo {title} {Generalized {$F$}-theorem and the
  {$\epsilon$} expansion},\ }\href {https://doi.org/10.1007/JHEP12(2015)155}
  {\bibfield  {journal} {\bibinfo  {journal} {J. High Energy Phys.}\ }\textbf
  {\bibinfo {volume} {2015}}\bibinfo  {number} { (12)},\ \bibinfo {pages}
  {155}}\BibitemShut {NoStop}%
\bibitem [{\citenamefont {Zhou}\ and\ \citenamefont
  {Ye}(2024)}]{zhou2024entanglement}%
  \BibitemOpen
\bibfield  {number} {  }\bibfield  {author} {\bibinfo {author} {\bibfnamefont
  {Y.}~\bibnamefont {Zhou}}\ and\ \bibinfo {author} {\bibfnamefont
  {P.}~\bibnamefont {Ye}},\ }\bibfield  {title} {\bibinfo {title} {Entanglement
  fractalization},\ }\href {https://doi.org/10.1103/PhysRevResearch.6.043145}
  {\bibfield  {journal} {\bibinfo  {journal} {Phys. Rev. Research}\ }\textbf
  {\bibinfo {volume} {6}},\ \bibinfo {pages} {043145} (\bibinfo {year}
  {2024})}\BibitemShut {NoStop}%
\bibitem [{\citenamefont {Kosior}\ and\ \citenamefont
  {Wang}(2026)}]{kosior2026entanglement}%
  \BibitemOpen
  \bibfield  {author} {\bibinfo {author} {\bibfnamefont {A.}~\bibnamefont
  {Kosior}}\ and\ \bibinfo {author} {\bibfnamefont {J.}~\bibnamefont {Wang}},\
  }\href@noop {} {\bibinfo {title} {Entanglement entropy of free fermions on
  random fractal lattices}} (\bibinfo {year} {2026}),\ \Eprint
  {https://arxiv.org/abs/2607.05611} {arXiv:2607.05611 [cond-mat.stat-mech]}
  \BibitemShut {NoStop}%
\bibitem [{\citenamefont {Peschel}(2003)}]{peschel2003calculation}%
  \BibitemOpen
  \bibfield  {author} {\bibinfo {author} {\bibfnamefont {I.}~\bibnamefont
  {Peschel}},\ }\bibfield  {title} {\bibinfo {title} {Calculation of reduced
  density matrices from correlation functions},\ }\href
  {https://doi.org/10.1088/0305-4470/36/14/101} {\bibfield  {journal} {\bibinfo
   {journal} {J. Phys. A: Math. Gen.}\ }\textbf {\bibinfo {volume} {36}},\
  \bibinfo {pages} {L205} (\bibinfo {year} {2003})}\BibitemShut {NoStop}%
\bibitem [{\citenamefont {Peschel}\ and\ \citenamefont
  {Eisler}(2009)}]{peschel2009reduced}%
  \BibitemOpen
  \bibfield  {author} {\bibinfo {author} {\bibfnamefont {I.}~\bibnamefont
  {Peschel}}\ and\ \bibinfo {author} {\bibfnamefont {V.}~\bibnamefont
  {Eisler}},\ }\bibfield  {title} {\bibinfo {title} {Reduced density matrices
  and entanglement entropy in free lattice models},\ }\href
  {https://doi.org/10.1088/1751-8113/42/50/504003} {\bibfield  {journal}
  {\bibinfo  {journal} {J. Phys. A: Math. Theor.}\ }\textbf {\bibinfo {volume}
  {42}},\ \bibinfo {pages} {504003} (\bibinfo {year} {2009})}\BibitemShut
  {NoStop}%
\bibitem [{\citenamefont {Zhou}(2022)}]{zhou2022floquet}%
  \BibitemOpen
  \bibfield  {author} {\bibinfo {author} {\bibfnamefont {L.}~\bibnamefont
  {Zhou}},\ }\bibfield  {title} {\bibinfo {title} {Entanglement spectrum and
  entropy in {Floquet} topological matter},\ }\href
  {https://doi.org/10.1103/PhysRevResearch.4.043164} {\bibfield  {journal}
  {\bibinfo  {journal} {Phys. Rev. Research}\ }\textbf {\bibinfo {volume}
  {4}},\ \bibinfo {pages} {043164} (\bibinfo {year} {2022})}\BibitemShut
  {NoStop}%
\bibitem [{\citenamefont {Beckermann}\ and\ \citenamefont
  {Townsend}(2017)}]{beckermann2017displacement}%
  \BibitemOpen
  \bibfield  {author} {\bibinfo {author} {\bibfnamefont {B.}~\bibnamefont
  {Beckermann}}\ and\ \bibinfo {author} {\bibfnamefont {A.}~\bibnamefont
  {Townsend}},\ }\bibfield  {title} {\bibinfo {title} {On the singular values
  of matrices with displacement structure},\ }\href
  {https://doi.org/10.1137/16M1096426} {\bibfield  {journal} {\bibinfo
  {journal} {SIAM J. Matrix Anal. Appl.}\ }\textbf {\bibinfo {volume} {38}},\
  \bibinfo {pages} {1227} (\bibinfo {year} {2017})}\BibitemShut {NoStop}%
\bibitem [{SM()}]{SM}%
  \BibitemOpen
  \href@noop {} {}\bibinfo {note} {See Supplementary Materials at [URL will be
  inserted by publisher] for the conditions and proofs of the entropy bounds
  and scaling laws in L1--L3, spectral-gap and zero-mode analyses, and
  numerical methods, fitting procedures, and additional tests on the
  {Sierpi{\'n}ski} carpet, {Vicsek} fractal, right-isosceles {Sierpi{\'n}ski}
  gasket lattice, and equilateral {Sierpi{\'n}ski} gasket lattice. The
  Supplemental Material also includes Refs.~\cite{lewis1996derivatives,
  bhatia1997matrix, klich2003elementary, klich2006lower, higham2008functions,
  brouwer2012spectra, benzi2013decay, trefethen2013approximation,
  condealonso2023schur, li2025geodesics}.}\BibitemShut {Stop}%
\bibitem [{\citenamefont {Inui}\ \emph {et~al.}(1994)\citenamefont {Inui},
  \citenamefont {Trugman},\ and\ \citenamefont {Abrahams}}]{inui1994unusual}%
  \BibitemOpen
  \bibfield  {author} {\bibinfo {author} {\bibfnamefont {M.}~\bibnamefont
  {Inui}}, \bibinfo {author} {\bibfnamefont {S.~A.}\ \bibnamefont {Trugman}},\
  and\ \bibinfo {author} {\bibfnamefont {E.}~\bibnamefont {Abrahams}},\
  }\bibfield  {title} {\bibinfo {title} {Unusual properties of midband states
  in systems with off-diagonal disorder},\ }\href
  {https://doi.org/10.1103/PhysRevB.49.3190} {\bibfield  {journal} {\bibinfo
  {journal} {Phys. Rev. B}\ }\textbf {\bibinfo {volume} {49}},\ \bibinfo
  {pages} {3190} (\bibinfo {year} {1994})}\BibitemShut {NoStop}%
\bibitem [{\citenamefont {Campostrini}\ \emph {et~al.}(2014)\citenamefont
  {Campostrini}, \citenamefont {Pelissetto},\ and\ \citenamefont
  {Vicari}}]{campostrini2014finite}%
  \BibitemOpen
  \bibfield  {author} {\bibinfo {author} {\bibfnamefont {M.}~\bibnamefont
  {Campostrini}}, \bibinfo {author} {\bibfnamefont {A.}~\bibnamefont
  {Pelissetto}},\ and\ \bibinfo {author} {\bibfnamefont {E.}~\bibnamefont
  {Vicari}},\ }\bibfield  {title} {\bibinfo {title} {Finite-size scaling at
  quantum transitions},\ }\href {https://doi.org/10.1103/PhysRevB.89.094516}
  {\bibfield  {journal} {\bibinfo  {journal} {Phys. Rev. B}\ }\textbf {\bibinfo
  {volume} {89}},\ \bibinfo {pages} {094516} (\bibinfo {year}
  {2014})}\BibitemShut {NoStop}%
\bibitem [{\citenamefont {Brzezi{\'n}ska}\ \emph {et~al.}(2018)\citenamefont
  {Brzezi{\'n}ska}, \citenamefont {Cook},\ and\ \citenamefont
  {Neupert}}]{brzezinska2018topology}%
  \BibitemOpen
  \bibfield  {author} {\bibinfo {author} {\bibfnamefont {M.}~\bibnamefont
  {Brzezi{\'n}ska}}, \bibinfo {author} {\bibfnamefont {A.~M.}\ \bibnamefont
  {Cook}},\ and\ \bibinfo {author} {\bibfnamefont {T.}~\bibnamefont
  {Neupert}},\ }\bibfield  {title} {\bibinfo {title} {Topology in the
  {Sierpi{\'n}ski}-{Hofstadter} problem},\ }\href
  {https://doi.org/10.1103/PhysRevB.98.205116} {\bibfield  {journal} {\bibinfo
  {journal} {Phys. Rev. B}\ }\textbf {\bibinfo {volume} {98}},\ \bibinfo
  {pages} {205116} (\bibinfo {year} {2018})}\BibitemShut {NoStop}%
\bibitem [{\citenamefont {Li}\ \emph {et~al.}(2026)\citenamefont {Li},
  \citenamefont {Tang}, \citenamefont {Liu},\ and\ \citenamefont
  {Hu}}]{li2026anderson}%
  \BibitemOpen
  \bibfield  {author} {\bibinfo {author} {\bibfnamefont {T.}~\bibnamefont
  {Li}}, \bibinfo {author} {\bibfnamefont {X.}~\bibnamefont {Tang}}, \bibinfo
  {author} {\bibfnamefont {S.}~\bibnamefont {Liu}},\ and\ \bibinfo {author}
  {\bibfnamefont {H.}~\bibnamefont {Hu}},\ }\href
  {https://doi.org/10.48550/arXiv.2605.17953} {\bibinfo {title} {Anderson
  transition and mobility edges in a family of {3D} fractal lattices}}
  (\bibinfo {year} {2026}),\ \Eprint {https://arxiv.org/abs/2605.17953}
  {arXiv:2605.17953 [cond-mat.dis-nn]} \BibitemShut {NoStop}%
\bibitem [{\citenamefont {Yang}\ \emph {et~al.}(2026)\citenamefont {Yang},
  \citenamefont {Chen}, \citenamefont {Zhao}, \citenamefont {Wu}, \citenamefont
  {Wang}, \citenamefont {Luo}, \citenamefont {Han}, \citenamefont {Chen},
  \citenamefont {Hu}, \citenamefont {Li}, \citenamefont {Li}, \citenamefont
  {Chen},\ and\ \citenamefont {Yang}}]{yang2026qiushi}%
  \BibitemOpen
  \bibfield  {author} {\bibinfo {author} {\bibfnamefont {S.}~\bibnamefont
  {Yang}}, \bibinfo {author} {\bibfnamefont {F.}~\bibnamefont {Chen}}, \bibinfo
  {author} {\bibfnamefont {R.}~\bibnamefont {Zhao}}, \bibinfo {author}
  {\bibfnamefont {J.}~\bibnamefont {Wu}}, \bibinfo {author} {\bibfnamefont
  {Y.}~\bibnamefont {Wang}}, \bibinfo {author} {\bibfnamefont {H.}~\bibnamefont
  {Luo}}, \bibinfo {author} {\bibfnamefont {N.}~\bibnamefont {Han}}, \bibinfo
  {author} {\bibfnamefont {Q.}~\bibnamefont {Chen}}, \bibinfo {author}
  {\bibfnamefont {Y.}~\bibnamefont {Hu}}, \bibinfo {author} {\bibfnamefont
  {W.}~\bibnamefont {Li}}, \bibinfo {author} {\bibfnamefont {M.}~\bibnamefont
  {Li}}, \bibinfo {author} {\bibfnamefont {H.}~\bibnamefont {Chen}},\ and\
  \bibinfo {author} {\bibfnamefont {Y.}~\bibnamefont {Yang}},\ }\href
  {https://doi.org/10.48550/arXiv.2604.27092} {\bibinfo {title} {End-to-end
  autonomous scientific discovery on a real optical platform}} (\bibinfo {year}
  {2026}),\ \Eprint {https://arxiv.org/abs/2604.27092} {arXiv:2604.27092}
  \BibitemShut {NoStop}%
\bibitem [{\citenamefont {Lewis}(1996)}]{lewis1996derivatives}%
  \BibitemOpen
  \bibfield  {author} {\bibinfo {author} {\bibfnamefont {A.~S.}\ \bibnamefont
  {Lewis}},\ }\bibfield  {title} {\bibinfo {title} {Derivatives of spectral
  functions},\ }\href {https://doi.org/10.1287/moor.21.3.576} {\bibfield
  {journal} {\bibinfo  {journal} {Math. Oper. Res.}\ }\textbf {\bibinfo
  {volume} {21}},\ \bibinfo {pages} {576} (\bibinfo {year} {1996})}\BibitemShut
  {NoStop}%
\bibitem [{\citenamefont {Bhatia}(1997)}]{bhatia1997matrix}%
  \BibitemOpen
  \bibfield  {author} {\bibinfo {author} {\bibfnamefont {R.}~\bibnamefont
  {Bhatia}},\ }\href {https://doi.org/10.1007/978-1-4612-0653-8} {\emph
  {\bibinfo {title} {Matrix Analysis}}},\ \bibinfo {series} {Graduate Texts in
  Mathematics}, Vol.\ \bibinfo {volume} {169}\ (\bibinfo  {publisher}
  {Springer},\ \bibinfo {address} {New York},\ \bibinfo {year}
  {1997})\BibitemShut {NoStop}%
\bibitem [{\citenamefont {Klich}(2003)}]{klich2003elementary}%
  \BibitemOpen
  \bibfield  {author} {\bibinfo {author} {\bibfnamefont {I.}~\bibnamefont
  {Klich}},\ }\bibfield  {title} {\bibinfo {title} {An elementary derivation of
  {Levitov}'s formula},\ }in\ \href
  {https://doi.org/10.1007/978-94-010-0089-5_19} {\emph {\bibinfo {booktitle}
  {Quantum Noise in Mesoscopic Physics}}},\ \bibinfo {series} {NATO Science
  Series}, Vol.~\bibinfo {volume} {97},\ \bibinfo {editor} {edited by\ \bibinfo
  {editor} {\bibfnamefont {Y.~V.}\ \bibnamefont {Nazarov}}}\ (\bibinfo
  {publisher} {Springer},\ \bibinfo {address} {Dordrecht},\ \bibinfo {year}
  {2003})\ pp.\ \bibinfo {pages} {397--402}\BibitemShut {NoStop}%
\bibitem [{\citenamefont {Klich}(2006)}]{klich2006lower}%
  \BibitemOpen
  \bibfield  {author} {\bibinfo {author} {\bibfnamefont {I.}~\bibnamefont
  {Klich}},\ }\bibfield  {title} {\bibinfo {title} {Lower entropy bounds and
  particle number fluctuations in a {Fermi} sea},\ }\href
  {https://doi.org/10.1088/0305-4470/39/4/L02} {\bibfield  {journal} {\bibinfo
  {journal} {J. Phys. A: Math. Gen.}\ }\textbf {\bibinfo {volume} {39}},\
  \bibinfo {pages} {L85} (\bibinfo {year} {2006})}\BibitemShut {NoStop}%
\bibitem [{\citenamefont {Higham}(2008)}]{higham2008functions}%
  \BibitemOpen
  \bibfield  {author} {\bibinfo {author} {\bibfnamefont {N.~J.}\ \bibnamefont
  {Higham}},\ }\href {https://doi.org/10.1137/1.9780898717778} {\emph {\bibinfo
  {title} {Functions of Matrices: Theory and Computation}}}\ (\bibinfo
  {publisher} {Society for Industrial and Applied Mathematics},\ \bibinfo
  {address} {Philadelphia},\ \bibinfo {year} {2008})\BibitemShut {NoStop}%
\bibitem [{\citenamefont {Brouwer}\ and\ \citenamefont
  {Haemers}(2012)}]{brouwer2012spectra}%
  \BibitemOpen
  \bibfield  {author} {\bibinfo {author} {\bibfnamefont {A.~E.}\ \bibnamefont
  {Brouwer}}\ and\ \bibinfo {author} {\bibfnamefont {W.~H.}\ \bibnamefont
  {Haemers}},\ }\href {https://doi.org/10.1007/978-1-4614-1939-6} {\emph
  {\bibinfo {title} {Spectra of Graphs}}},\ Universitext\ (\bibinfo
  {publisher} {Springer},\ \bibinfo {address} {New York},\ \bibinfo {year}
  {2012})\BibitemShut {NoStop}%
\bibitem [{\citenamefont {Benzi}\ \emph {et~al.}(2013)\citenamefont {Benzi},
  \citenamefont {Boito},\ and\ \citenamefont {Razouk}}]{benzi2013decay}%
  \BibitemOpen
  \bibfield  {author} {\bibinfo {author} {\bibfnamefont {M.}~\bibnamefont
  {Benzi}}, \bibinfo {author} {\bibfnamefont {P.}~\bibnamefont {Boito}},\ and\
  \bibinfo {author} {\bibfnamefont {N.}~\bibnamefont {Razouk}},\ }\bibfield
  {title} {\bibinfo {title} {Decay properties of spectral projectors with
  applications to electronic structure},\ }\href
  {https://doi.org/10.1137/100814019} {\bibfield  {journal} {\bibinfo
  {journal} {SIAM Rev.}\ }\textbf {\bibinfo {volume} {55}},\ \bibinfo {pages}
  {3} (\bibinfo {year} {2013})}\BibitemShut {NoStop}%
\bibitem [{\citenamefont {Trefethen}(2013)}]{trefethen2013approximation}%
  \BibitemOpen
  \bibfield  {author} {\bibinfo {author} {\bibfnamefont {L.~N.}\ \bibnamefont
  {Trefethen}},\ }\href {https://www.chebfun.org/ATAP/} {\emph {\bibinfo
  {title} {Approximation Theory and Approximation Practice}}}\ (\bibinfo
  {publisher} {Society for Industrial and Applied Mathematics},\ \bibinfo
  {address} {Philadelphia},\ \bibinfo {year} {2013})\BibitemShut {NoStop}%
\bibitem [{\citenamefont {Conde-Alonso}\ \emph {et~al.}(2023)\citenamefont
  {Conde-Alonso}, \citenamefont {Gonz{\'a}lez-P{\'e}rez}, \citenamefont
  {Parcet},\ and\ \citenamefont {Tablate}}]{condealonso2023schur}%
  \BibitemOpen
  \bibfield  {author} {\bibinfo {author} {\bibfnamefont {J.~M.}\ \bibnamefont
  {Conde-Alonso}}, \bibinfo {author} {\bibfnamefont {A.~M.}\ \bibnamefont
  {Gonz{\'a}lez-P{\'e}rez}}, \bibinfo {author} {\bibfnamefont {J.}~\bibnamefont
  {Parcet}},\ and\ \bibinfo {author} {\bibfnamefont {E.}~\bibnamefont
  {Tablate}},\ }\bibfield  {title} {\bibinfo {title} {Schur multipliers in
  {Schatten-von Neumann} classes},\ }\href
  {https://doi.org/10.4007/annals.2023.198.3.5} {\bibfield  {journal} {\bibinfo
   {journal} {Ann. Math.}\ }\textbf {\bibinfo {volume} {198}},\ \bibinfo
  {pages} {1229} (\bibinfo {year} {2023})}\BibitemShut {NoStop}%
\bibitem [{\citenamefont {Li}\ \emph {et~al.}(2025)\citenamefont {Li},
  \citenamefont {Liang},\ and\ \citenamefont {Xue}}]{li2025geodesics}%
  \BibitemOpen
  \bibfield  {author} {\bibinfo {author} {\bibfnamefont {X.}~\bibnamefont
  {Li}}, \bibinfo {author} {\bibfnamefont {X.}~\bibnamefont {Liang}},\ and\
  \bibinfo {author} {\bibfnamefont {Y.}~\bibnamefont {Xue}},\ }\bibfield
  {title} {\bibinfo {title} {Geodesics of right isosceles {Sierpi{\'n}ski}
  gaskets and their relatives},\ }\href
  {https://doi.org/10.1016/j.chaos.2024.115937} {\bibfield  {journal} {\bibinfo
   {journal} {Chaos Solitons Fractals}\ }\textbf {\bibinfo {volume} {192}},\
  \bibinfo {pages} {115937} (\bibinfo {year} {2025})}\BibitemShut {NoStop}%
\end{thebibliography}
\end{document}